\documentclass[12pt,thmsa]{article}
\usepackage{amsmath}
\usepackage{graphicx}
\usepackage{amsfonts}
\usepackage{amssymb}

\begin{document}
\textbf{The proof that the standard transformations of }$E$\textbf{ and }$B$\textbf{\ are}

\textbf{not\ the\ Lorentz transformations. Clifford\ algebra formalism}\bigskip

\qquad Tomislav Ivezi\'{c}

\qquad\textit{Ru\mbox
{\it{d}\hspace{-.15em}\rule[1.25ex]{.2em}{.04ex}\hspace{-.05em}}er Bo\v
{s}kovi\'{c} Institute, P.O.B. 180, 10002 Zagreb, Croatia}

\textit{\qquad ivezic@irb.hr\bigskip}

In this paper it is exactly proved by using the Clifford algebra formalism
that the standard transformations of the three-dimensional (3D) vectors of the
electric and magnetic fields $\mathbf{E}$ and $\mathbf{B}$ \emph{are not} the
Lorentz transformations of well-defined quantities from the 4D spacetime but
the 'apparent' transformations of the 3D quantities. Thence the usual Maxwell
equations with the 3D $\mathbf{E}$ and $\mathbf{B}$ \emph{are not} in
agreement with special relativity. The 1-vectors $E$ and $B,$ as well-defined
4D quantities, are introduced instead of ill-defined 3D $\mathbf{E}$ and
$\mathbf{B.}$ \bigskip

\noindent PACS: 03.30.+p \bigskip

\noindent\textit{Keywords}: standard transformations of $\mathbf{E}$,
$\mathbf{B;}$ special relativity; Clifford algebra\bigskip\bigskip

\noindent\textbf{1. Introduction} \bigskip

It is generally accepted by physics community that there is an agreement
between the classical electromagnetism and the special relativity (SR). Such
opinion is prevailing in physics already from the Einstein first paper on SR
[1]. The standard transformations of the 3D vectors of the electric and
magnetic fields, $\mathbf{E}$ and $\mathbf{B}$ respectively, are first derived
by Lorentz [2] and independently by Einstein [1], and subsequently quoted in
almost every textbook and paper on relativistic electrodynamics. They are
considered to be the Lorentz transformations (LT) of these vectors, see, e.g.,
[1-3]. The same opinion holds in all usual Clifford algebra formulations of
the classical electromagnetism, e.g., the formulations with Clifford
multivectors [4-6]. The usual Maxwell equations with the 3D vectors
$\mathbf{E}$ and $\mathbf{B}$ are assumed to be physically equivalent to the
field equations expressed in terms of the Faraday bivector field $F$ in the
Clifford algebra formalism (or the electromagnetic field tensor $F^{ab}$ in
the tensor formalism). In this paper it will be exactly proved that the above
mentioned standard transformations of $\mathbf{E}$ and $\mathbf{B}$ (see eq.
(\ref{sk1}) below) \emph{are not} relativistically correct transformations in
the 4D spacetime; they are not the LT of the 3D $\mathbf{E}$ and $\mathbf{B.}$
Consequently the usual Maxwell equations with $\mathbf{E}$ and $\mathbf{B}$
and the field equations with the $F$ field \emph{are not} physically
equivalent. The correct LT (the active ones) of the electric and magnetic
fields are given by the relations (\ref{nle}) and (\ref{nlb}) (or (\ref{eh})
and (\ref{Be})) below. In the Clifford algebra formalism (as in the tensor
formalism) one deals either with 4D quantities that are defined without
reference frames, e.g., Clifford multivector $F$ (the abstract tensor $F^{ab}%
$) or, when some basis has been introduced, with coordinate-based geometric
quantity that comprises both components and a basis. The SR that exclusively
deals with quantities defined without reference frames or, equivalently, with
coordinate-based geometric quantities, can be called the invariant SR. The
reason for this name is that upon the passive LT any coordinate-based
geometric quantity remains unchanged. The invariance of some 4D
coordinate-based geometric quantity upon the passive LT reflects the fact that
such mathematical, invariant, geometric 4D quantity represents \emph{the same
physical object} for relatively moving observers. \emph{It is taken in the
invariant SR that such 4D geometric quantities are well-defined not only
mathematically but also experimentally, as measurable quantities with real
physical meaning. Thus they do have an independent physical reality. }The
invariant SR is discussed in [7] in the Clifford algebra formalism and in
[8,9] in the tensor formalism. It is explicitly shown in [9] that the true
agreement with experiments that test SR exists when the theory deals with
well-defined 4D quantities, i.e., the quantities that are invariant upon the
passive LT. The usual standard transformations of the electric and magnetic
fields, the transformations (\ref{ce}), (\ref{B}) and (\ref{sk1}) (or
(\ref{es}) and (\ref{bes})) below are typical examples of the `apparent'
transformations that are first discussed in [10] and [11]. The `apparent'
transformations of the spatial distances (the Lorentz contraction) and the
temporal distances (the dilatation of time) are elaborated in detail in [8]
and [9] (see also [12]), and in [8] I have also discussed in the tensor
formalism the `apparent' transformations of the 3D vectors $\mathbf{E}$ and
$\mathbf{B}$. The `apparent' transformations relate, in fact, the quantities
from '3+1' space \emph{and} time (spatial and temporal distances and 3D
vectors $\mathbf{E}$ and $\mathbf{B}$) and not well-defined 4D quantities.
But, in contrast to the LT of well-defined 4D quantities, \emph{the `apparent'
transformations do not refer to the same physical object} for relatively
moving observers. In this paper it will be also shown that in the 4D spacetime
the well-defined 4D quantities, the 1-vectors of the electric and magnetic
fields $E$ and $B$ (see (\ref{itf})) in the Clifford algebra formalism (as in
[7]), have to be introduced instead of ill-defined 3D vectors $\mathbf{E}$ and
$\mathbf{B.}$ The same proof is already presented in the tensor formalism in
[13]. \bigskip\medskip

\noindent\textbf{2. The}\ $\gamma_{0}$ - \textbf{split and\ the usual
expressions for} $\mathbf{E}$\ \textbf{and}\ $\mathbf{B}$\ \textbf{in
the}\ $\gamma_{0}$ - \textbf{frame} \bigskip

\noindent\textit{2.1. A brief summary of geometric algebra} \medskip

First we provide a brief summary of Clifford algebra with multivectors (see,
e.g., $\left[  4-6\right]  $). We write Clifford vectors in lower case ($a$)
and general multivectors (Clifford aggregate) in upper case ($A$). The space
of multivectors is graded and multivectors containing elements of a single
grade, $r$, are termed homogeneous and written $A_{r}.$ The geometric
(Clifford) product is written by simply juxtaposing multivectors $AB$. A basic
operation on multivectors is the degree projection $\left\langle
A\right\rangle _{r}$ which selects from the multivector $A$ its $r-$ vector
part ($0=$ scalar, $1=$ vector, $2=$ bivector ....). We write the scalar
(grade-$0$) part simply as $\left\langle A\right\rangle .$ The geometric
product of a grade-$r$ multivector $A_{r}$ with a grade-$s$ multivector
$B_{s}$ decomposes into $A_{r}B_{s}=\left\langle AB\right\rangle
_{\ r+s}+\left\langle AB\right\rangle _{\ r+s-2}...+\left\langle
AB\right\rangle _{\ \left|  r-s\right|  }.$ The inner and outer (or exterior)
products are the lowest-grade and the highest-grade terms respectively of the
above series $A_{r}\cdot B_{s}\equiv\left\langle AB\right\rangle _{\ \left|
r-s\right|  },$ and $A_{r}\wedge B_{s}\equiv\left\langle AB\right\rangle
_{\ r+s}.$ For vectors $a$ and $b$ we have $ab=a\cdot b+a\wedge b,$ where
$a\cdot b\equiv(1/2)(ab+ba),$ and $a\wedge b\equiv(1/2)(ab-ba).$ Reversion is
an invariant kind of conjugation, which is defined by $\widetilde
{AB}=\widetilde{B}\widetilde{A},$ $\widetilde{a}=a,$ for any vector $a$, and
it reverses the order of vectors in any given expression. Any multivector $A$
is a geometric 4D quantity defined without reference frame. When some basis
has been introduced $A$ can be written as a coordinate-based geometric
quantity comprising both components and a basis. Usually, e.g., $\left[
4-6\right]  $, one introduces the standard basis. The generators of the
spacetime algebra are taken to be four basis vectors $\left\{  \gamma_{\mu
}\right\}  ,\mu=0...3,$ satisfying $\gamma_{\mu}\cdot\gamma_{\nu}=\eta_{\mu
\nu}=diag(+---).$ This basis is a right-handed orthonormal frame of vectors in
the Minkowski spacetime $M^{4}$ with $\gamma_{0}$ in the forward light cone.
The $\gamma_{k}$ ($k=1,2,3$) are spacelike vectors. This algebra is often
called the Dirac algebra $D$ and the elements of $D$ are called $d-$numbers.
The $\gamma_{\mu}$ generate by multiplication a complete basis, the standard
basis, for spacetime algebra: $1,\gamma_{\mu},\gamma_{\mu}\wedge\gamma_{\nu
},\gamma_{\mu}\gamma_{5,}\gamma_{5}$ ($16$ independent elements). $\gamma_{5}$
is the pseudoscalar for the frame $\left\{  \gamma_{\mu}\right\}  .$

We remark that the standard basis corresponds, in fact, to the specific system
of coordinates, i.e., to Einstein's system of coordinates. In the Einstein
system of coordinates the Einstein synchronization $\left[  1\right]  $ of
distant clocks and Cartesian space coordinates $x^{i}$ are used in the chosen
inertial frame of reference. However different systems of coordinates of an
inertial frame of reference are allowed and they are all equivalent in the
description of physical phenomena. For example, in $\left[  8\right]  $ two
very different, but completely equivalent systems of coordinates, the Einstein
system of coordinates and ''radio'' (''r'') system of coordinates, are exposed
and exploited throughout the paper. The coordinate-based geometric quantities
representing some 4D physical quantity in different relatively moving inertial
frames of reference, or in different systems of coordinates in the chosen
inertial frame of reference, are all mathematically equal and thus they are
\emph{the same quantity }for different observers, or in different systems of
coordinates. Then, e.g., the position 1-vector $x$ (a geometric quantity) can
be decomposed in the $S$ and $S^{\prime}$ frames and in the standard basis
$\left\{  \gamma_{\mu}\right\}  $ as $x=x^{\mu}\gamma_{\mu}=x^{\prime\mu
}\gamma_{\mu}^{\prime}.$ The primed quantities are the Lorentz transforms of
the unprimed ones. In such interpretation the LT are considered as passive
transformations; both the components and the basis vectors are transformed but
the whole geometric quantity remains unchanged. Thus we see that \emph{under
the passive LT a well-defined quantity on the 4D spacetime, i.e., a
coordinate-based geometric quantity, is an invariant quantity.} As already
said in the Introduction the SR that exclusively deals with such quantities
defined without reference frames or, equivalently, with coordinate-based
geometric quantities, is called the invariant SR and it is considered in the
tensor formalism in [8,9].

In the usual Clifford algebra formalism $\left[  4-6\right]  $ instead of
working only with such \emph{observer independent quantities} one introduces a
space-time split and the relative vectors. By singling out a particular
time-like direction $\gamma_{0}$ we can get a unique mapping of spacetime into
the even subalgebra of spacetime algebra. For each event $x$ this mapping is
specified by $x\gamma_{0}=ct+\mathbf{x,\quad} ct=x\cdot\gamma_{0}%
,\ \mathbf{x}=x\wedge\gamma_{0}$. The set of all position vectors $\mathbf{x}$
is the 3D position space of the observer $\gamma_{0}$ and it is designated by
$P^{3}$. The elements of $P^{3}$ are called \textit{the relative vectors}
(relative to $\gamma_{0})$ and they will be designated in boldface. The
explicit appearance of $\gamma_{0}$ implies that \emph{the space-time split is
observer dependent}. If we consider the position 1-vector $x$ in another
relatively moving inertial frame of reference $S^{\prime}$ (characterized by
$\gamma_{0}^{\prime}$) then the space-time split in $S^{\prime}$ and in the
Einstein system of coordinates is $x\gamma_{0}^{\prime}=ct^{\prime}%
+\mathbf{x}^{\prime}$\textbf{.} This $x\gamma_{0}^{\prime}$ is not obtained by
the LT from $x\gamma_{0}.$ (The hypersurface $t^{\prime}=const.$ is not
connected in any way with the hypersurface $t=const.$) Thence the spatial and
the temporal components ($\mathbf{x}$, $t$) of some geometric 4D quantity
($x$) (and thus the relative vectors as well) are not physically well-defined
quantities. Only their union is physically well-defined quantity in the 4D
spacetime from the invariant SR viewpoint.\bigskip

\noindent\textit{2.2. The\ }$\gamma_{0}$\textit{ -
split\ and\ the\ usual\ expressions for }$E$\textit{\ and }$B$\textit{\ in the
}$\gamma_{0}$\textit{ - frame}\textbf{\medskip}

Let us now see how the space-time split is introduced in the usual Clifford
algebra formalism [4,5] of the electromagnetism. The bivector field $F$ is
expressed in terms of the sum of a relative vector $\mathbf{E}$ and a relative
bivector $\gamma_{5}\mathbf{B}$ by making a space-time split in the
$\gamma_{0}$ - frame
\begin{align}
F &  =\mathbf{E}_{H}+c\gamma_{5}\mathbf{B}_{H}\mathbf{,\quad E}_{H}%
=(F\cdot\gamma_{0})\gamma_{0}=(1/2)(F-\gamma_{0}F\gamma_{0}),\nonumber\\
\gamma_{5}\mathbf{B}_{H} &  =(1/c)(F\wedge\gamma_{0})\gamma_{0}%
=(1/2c)(F+\gamma_{0}F\gamma_{0}).\label{FB}%
\end{align}
(The subscript 'H' is for - Hestenes.) Both $\mathbf{E}_{H}$ and
$\mathbf{B}_{H}$ are, in fact, bivectors. Similarly in [6] $F$ is decomposed
in terms of 1-vector $\mathbf{E}_{J}$ and a bivector $\mathbf{B}_{J}$ (the
subscript 'J' is for - Jancewicz) as
\begin{equation}
F=\gamma_{0}\wedge\mathbf{E}_{J}-c\mathbf{B}_{J},\quad\mathbf{E}_{J}%
=F\cdot\gamma_{0},\ \mathbf{B}_{J}=-(1/c)(F\wedge\gamma_{0})\gamma
_{0}.\label{J}%
\end{equation}
Instead of to use $\mathbf{E}_{H},$ $\mathbf{B}_{H}$ or $\mathbf{E}_{J},$
$\mathbf{B}_{J}$ we shall mainly deal (except in Sec. 3.3.) with simpler but
completely equivalent expressions in the $\gamma_{0}$ - frame, i.e., with
1-vectors that will be denoted as $E_{f}$ and $B_{f}.$ Then
\begin{align}
F &  =E_{f}\wedge\gamma_{0}+c(\gamma_{5}B_{f})\cdot\gamma_{0},\nonumber\\
E_{f} &  =F\cdot\gamma_{0},\ B_{f}=-(1/c)\gamma_{5}(F\wedge\gamma
_{0}).\label{ebg}%
\end{align}
All these quantities can be written as coordinate-based geometric quantities
in the standard basis $\left\{  \gamma_{\mu}\right\}  .$ Thus
\begin{equation}
F=(1/2)F^{\mu\nu}\gamma_{\mu}\wedge\gamma_{\nu}=F^{0k}\gamma_{0}\wedge
\gamma_{k}+(1/2)F^{kl}\gamma_{k}\wedge\gamma_{l},\quad k,l=1,2,3.\label{EF}%
\end{equation}%
\begin{align}
E_{f} &  =E_{f}^{\mu}\gamma_{\mu}=0\gamma_{0}+F^{k0}\gamma_{k},\nonumber\\
B_{f} &  =B_{f}^{\mu}\gamma_{\mu}=0\gamma_{0}+(-1/2c)\varepsilon^{0kli}%
F_{kl}\gamma_{i}.\label{gnl}%
\end{align}
We see from (\ref{EF}) and (\ref{gnl}) that the components of $F$ in the
$\left\{  \gamma_{\mu}\right\}  $ basis (i.e., in the Einstein system of
coordinates) give rise to the tensor (components)$\;F^{\mu\nu}=\gamma^{\nu
}\cdot(\gamma^{\mu}\cdot F)=(\gamma^{\nu}\wedge\gamma^{\mu})\cdot F,$ which,
written out as a matrix, has entries
\begin{equation}
E_{f}^{i}=F^{i0},\quad B_{f}^{i}=(-1/2c)\varepsilon^{0kli}F_{kl}.\label{sko}%
\end{equation}
The relation (\ref{sko}) is nothing else than the standard identification of
the components $F^{\mu\nu}$ with the components of the 3D vectors $\mathbf{E}$
and $\mathbf{B,}$ see, e.g., [3]. It is worth noting that \emph{all
expressions with} $\gamma_{0}$ (\ref{ebg}) \emph{actually refer to the 3D
subspace orthogonal to the specific timelike direction }$\gamma_{0}.$ Really
it can be easily checked that $E_{f}\cdot\gamma_{0}=B_{f}\cdot\gamma_{0}=0,$
which means that they are orthogonal to $\gamma_{0};$ $E_{f}$ \emph{and}
$B_{f}$ \emph{do not have the temporal components} $E_{f}^{0}=B_{f}^{0}=0$.
These results (\ref{sko}) are quoted in numerous textbooks and papers treating
relativistic electrodynamics, see, e.g., [3]. Actually in the usual covariant
approaches, e.g., [3], one forgets about temporal components $E_{f}^{0}$ and
$B_{f}^{0}$ and simply makes the identification of six independent components
of $F^{\mu\nu}$ with three components $E_{f}^{i}$ and three components
$B_{f}^{i}$ according to (\ref{sko}). Since in SR we work with the 4D
spacetime the mapping between some components of $F^{\mu\nu}$ and the
components of the 3D vectors $\mathbf{E}$ and $\mathbf{B}$ is mathematically
better founded by the relations (\ref{gnl}) than by their simple
identification. Note that the whole procedure is made in an inertial frame of
reference with the Einstein system of coordinates. In another system of
coordinates that is different than the Einstein system of coordinates, e.g.,
differing in the chosen synchronization (as it is the 'r' synchronization
considered in [8]), the identification of $E_{f}^{i}$ with $F^{i0},$ as in
(\ref{sko}) (and also for $B_{f}^{i}$), is impossible and meaningless.\bigskip\medskip

\noindent\textbf{3}. \textbf{The\ proof\ that
the\ standard\ transformations\ of\ }$\mathbf{E}$\textbf{\ and\ }$\mathbf{B}%
$\ \textbf{are\ not\ the\ LT\bigskip}

\noindent\textit{3.1. The active LT of the electric and magnetic
fields}\textbf{ \medskip}

Let us now explicitly show that the usual transformations of the 3D
$\mathbf{E}$ and $\mathbf{B}$ are not relativistically\ correct, i.e., they
are not the LT of quantities that are well-defined on the 4D spacetime. First
we find the correct expressions for the LT (the active ones) of $E_{f}$ and
$B_{f}.$ In the usual Clifford algebra formalism, e.g., $\left[  4-6\right]
$, the LT are considered as active transformations; the components of, e.g.,
some 1-vector relative to a given inertial frame of reference (with the
standard basis $\left\{  \gamma_{\mu}\right\}  $) are transformed into the
components of a new 1-vector relative to the same frame (the basis $\left\{
\gamma_{\mu}\right\}  $ is not changed). Furthermore the LT are described with
rotors $R,$ $R\widetilde{R}=1,$ in the usual way as $p\rightarrow p^{\prime
}=Rp\widetilde{R}=p_{\mu}^{\prime}\gamma^{\mu}.$ But every rotor in spacetime
can be written in terms of a bivector as $R=e^{\theta/2}.$ For boosts in
arbitrary direction
\begin{equation}
R=e^{\theta/2}=(1+\gamma+\gamma\beta\gamma_{0}n)/(2(1+\gamma))^{1/2},
\label{err}%
\end{equation}
$\theta=\alpha\gamma_{0}n,$ $\beta$ is the scalar velocity in units of $c$,
$\gamma=(1-\beta^{2})^{-1/2}$, or in terms of an `angle' $\alpha$ we have
$\tanh\alpha=\beta,$ $\cosh\alpha=\gamma,$ $\sinh\alpha=\beta\gamma,$ and $n$
is not the basis vector but any unit space-like vector orthogonal to
$\gamma_{0};$ $e^{\theta}=\cosh\alpha+\gamma_{0}n\sinh\alpha.$ One can also
express the relationship between the two relatively moving frames $S$ and
$S^{\prime}$ in terms of rotor as $\gamma_{\mu}^{\prime}=R\gamma_{\mu
}\widetilde{R}.$ For boosts in the direction $\gamma_{1}$ the rotor $R$ is
given by the relation (\ref{err}) with $\gamma_{1}$ replacing $n$ (all in the
standard basis $\left\{  \gamma_{\mu}\right\}  $). Then using (\ref{gnl}) the
transformed $E_{f}^{\prime}$ can be written as
\begin{align}
E_{f}^{\prime}  &  =R(F\cdot\gamma_{0})\widetilde{R}=R(F^{k0}\gamma
_{k})\widetilde{R}=E_{f}^{\prime\mu}\gamma_{\mu}=\nonumber\\
&  =-\beta\gamma E_{f}^{1}\gamma_{0}+\gamma E_{f}^{1}\gamma_{1}+E_{f}%
^{2}\gamma_{2}+E_{f}^{3}\gamma_{3}, \label{nle}%
\end{align}
what is the usual form for the active LT of the 1-vector $E_{f}=E_{f}^{\mu
}\gamma_{\mu}$. Similarly we find for $B_{f}^{\prime}$
\begin{align}
B_{f}^{\prime}  &  =R\left[  -(1/c)\gamma_{5}(F\wedge\gamma_{0})\right]
\widetilde{R}=R\left[  (-1/2c)\varepsilon^{0kli}F_{kl}\gamma_{i}\right]
\widetilde{R}=\nonumber\\
&  =B_{f}^{\prime\mu}\gamma_{\mu}=-\beta\gamma B_{f}^{1}\gamma_{0}+\gamma
B_{f}^{1}\gamma_{1}+B_{f}^{2}\gamma_{2}+B_{f}^{3}\gamma_{3}, \label{nlb}%
\end{align}
what is the familiar form for the active LT of the 1-vector $B_{f}=B_{f}^{\mu
}\gamma_{\mu}$. It is important to note that $E_{f}^{\prime}$ \emph{and}
$B_{f}^{\prime}$ \emph{are not orthogonal to} $\gamma_{0},$ i.e., \emph{they}
\emph{do have} \emph{the temporal components} $\neq0.$ They do not belong to
the same 3D subspace as $E_{f}$ and $B_{f},$ but they are in the 4D spacetime
spanned by the whole standard basis $\left\{  \gamma_{\mu}\right\}  $. The
relations (\ref{nle}) and (\ref{nlb}) imply that the space-time split in the
$\gamma_{0}$ - system is not possible for the transformed $F^{\prime
}=RF\widetilde{R}$, i.e., $F^{\prime}$ cannot be decomposed into
$E_{f}^{\prime}$ and $B_{f}^{\prime}$ as $F$ is decomposed in the relation
(\ref{ebg}), $F^{\prime}\neq E_{f}^{\prime}\wedge\gamma_{0}+c(\gamma_{5}%
B_{f}^{\prime})\cdot\gamma_{0}.$ Notice, what is very important, that
\emph{the components} $E_{f}^{\mu}$ ($B_{f}^{\mu}$) from (\ref{gnl})
\emph{transform upon the active LT again to the components} $E_{f}^{\prime\mu
}$ ($B_{f}^{\prime\mu}$) from (\ref{nle}) ((\ref{nlb})); \emph{there is no
mixing of components}. \emph{Thus} \emph{by the active LT} $E_{f}$
\emph{transforms to} $E_{f}^{\prime}$ \emph{and} $B_{f}$ \emph{to }%
$B_{f}^{\prime}.$ Actually, as we said, this is the way in which every
1-vector transforms upon the active LT.\bigskip

\noindent\textit{3.2. The\ standard transformations of the electric and
magnetic fields}\textbf{ }\medskip

However \emph{the standard transformations for} $E_{st}^{\prime}$ \emph{and}
$B_{st}^{\prime}$ (the subscript - st - is for - standard)\emph{\ are derived
wrongly assuming that the quantities obtained by the active LT of} $E_{f}$
\emph{and} $B_{f}$ \emph{are again in the 3D subspace of the} $\gamma_{0}$
\emph{-} \emph{observer}. This means that it is wrongly assumed in all
standard derivations, e.g., in the Clifford algebra formalism [4-6] (and in
the tensor formalism [3] as well), that one can again perform the same
identification of the transformed components $F^{\prime\mu\nu}$ with the
components of the 3D $\mathbf{E}^{\prime}$ and $\mathbf{B}^{\prime}\mathbf{.}$
Thus it is taken in standard derivations that for the transformed
$E_{st}^{\prime}$ and $B_{st}^{\prime}$ again hold $E_{st}^{\prime0}%
=B_{st}^{\prime0}=0,$ i.e., that $E_{st}^{\prime}\cdot\gamma_{0}%
=B_{st}^{\prime}\cdot\gamma_{0}=0$ as for $E_{f}$ and $B_{f}.$ Thence, in
contrast to the correct LT of $E_{f}$ and $B_{f},$ (\ref{nle}) and (\ref{nlb})
respectively, it is taken in standard derivations that
\begin{align}
E_{st}^{\prime}  &  =(RF\widetilde{R})\cdot\gamma_{0}=F^{\prime}\cdot
\gamma_{0}=F^{\prime k0}\gamma_{k}=E_{st}^{\prime k}\gamma_{k}=\nonumber\\
&  =E_{f}^{1}\gamma_{1}+(\gamma E_{f}^{2}-\beta\gamma cB_{f}^{3})\gamma
_{2}+(\gamma E_{f}^{3}+\beta\gamma cB_{f}^{2})\gamma_{3}, \label{ce}%
\end{align}
where $F^{\prime}=RF\widetilde{R}$. Similarly we find for $B_{st}^{\prime}$
\begin{align}
B_{st}^{\prime}  &  =-(1/c)\gamma_{5}(F^{\prime}\wedge\gamma_{0}%
)=-(1/2c)\varepsilon^{0kli}F_{kl}^{\prime}\gamma_{i}=B_{st}^{\prime i}%
\gamma_{i}=\nonumber\\
&  B_{f}^{1}\gamma_{1}+(\gamma B_{f}^{2}+\beta\gamma E_{f}^{3}/c)\gamma
_{2}+(\gamma B_{f}^{3}-\beta\gamma E_{f}^{2}/c)\gamma_{3}. \label{B}%
\end{align}
From \emph{the relativistically incorrect transformations} (\ref{ce}) and
(\ref{B}) one simply finds the transformations of the spatial components
$E_{st}^{\prime i}$ and $B_{st}^{\prime i}$
\begin{equation}
E_{st}^{\prime i}=F^{\prime i0},\quad B_{st}^{\prime i}=(-1/2c)\varepsilon
^{0kli}F_{kl}^{\prime}. \label{sk1}%
\end{equation}
As can be seen from (\ref{ce}), (\ref{B}) and (\ref{sk1}) \emph{the
transformations for} $E_{st.}^{\prime i}$ \emph{and} $B_{st.}^{\prime i}$
\emph{are exactly the standard transformations of components of the 3D
vectors} $\mathbf{E}$ \emph{and} $\mathbf{B}$ that are quoted in almost every
textbook and paper on relativistic electrodynamics including [1] and [3].
These relations are explicitly derived and given in the Clifford algebra
formalism, e.g., in [4], Space-Time Algebra (eq. (18.22)), New Foundations for
Classical Mechanics (Ch. 9 eqs. (3.51a,b)) and in [6] (Ch. 7 eqs. (20a,b)).
Notice that, in contrast to the active LT (\ref{nle}) and (\ref{nlb}),
\emph{according to the standard transformations} (\ref{ce}) \emph{and
}(\ref{B}) (\emph{i.e.}, (\ref{sk1})) \emph{the transformed components}
$E_{st}^{\prime i}$ \emph{are expressed by the mixture of components}
$E_{f}^{i}$ \emph{and} $B_{f}^{i},$ \emph{and the same holds for}
$B_{st}^{\prime i}$. In all previous treatments of SR, e.g., [4-6] (and [1-3])
the transformations for $E_{st.}^{\prime i}$ and $B_{st.}^{\prime i}$ are
considered to be the LT of the 3D electric and magnetic fields. However our
analysis shows that the transformations for $E_{st.}^{\prime i}$ and
$B_{st.}^{\prime i}$ (\ref{sk1}) are derived from \emph{the relativistically
incorrect transformations} (\ref{ce}) and (\ref{B}), which are not the LT; the
LT are given by the relations (\ref{nle}) and (\ref{nlb}).

The same results can be obtained with the passive LT, either by using the
expression for the LT that is independent of the chosen system of coordinates
(such one as in [7]), or by using the standard expressions for the LT in the
Einstein system of coordinates from [3]. The passive LT transform always the
whole 4D quantity, basis and components, leaving the whole quantity unchanged.
Thus under the passive LT the field bivector $F$ as well-defined 4D quantity
remains unchanged, i.e., $F=(1/2)F^{\mu\nu}\gamma_{\mu}\wedge\gamma_{\nu
}=(1/2)F^{\prime\mu\nu}\gamma_{\mu}^{\prime}\wedge\gamma_{\nu}^{\prime}$ (all
primed quantities are the Lorentz transforms of the unprimed ones). In the
same way it holds that, e.g., $E_{f}^{\mu}\gamma_{\mu}=E_{f}^{\prime\mu}%
\gamma_{\mu}^{\prime}$. The invariance of some 4D coordinate-based geometric
quantity upon the passive LT is the crucial requirement that must be satisfied
by any well-defined 4D quantity. It reflects the fact that such mathematical,
invariant, geometric 4D quantity represents \emph{the same physical object}
for relatively moving observers. The use of coordinate-based geometric
quantities enables us to have clearly and correctly defined the concept of
sameness of a physical system for different observers. Thus \emph{such
quantity that does not change upon the passive LT does have an independent
physical reality, both theoretically and experimentally. }

However it can be easily shown that $E_{f}^{\mu}\gamma_{\mu}\neq
E_{st}^{\prime\mu}\gamma_{\mu}^{\prime}.$ This means that, e.g., $E_{f}^{\mu
}\gamma_{\mu}$ and $E_{st.}^{\prime\mu}\gamma_{\mu}^{\prime}$ \emph{are not
the same quantity for observers in} $S$ \emph{and} $S^{\prime}.$ As far as
relativity is concerned the quantities, e.g., $E_{f}^{\mu}\gamma_{\mu}$ and
$E_{st.}^{\prime\mu}\gamma_{\mu}^{\prime},$ are not related to one another.
Their identification is the typical case of \emph{mistaken identity}. The fact
that they are measured by two observers ($\gamma_{0}$ - and $\gamma
_{0}^{\prime}$ - observers) does not mean that relativity has something to do
with the problem. The reason is that observers in the $\gamma_{0}$ - system
and in the $\gamma_{0}^{\prime}$ - system are not looking at the same physical
object but at two different objects. \emph{Every observer makes measurement on
its own object and such measurements are not related by the LT.} Thus from the
point of view of the SR the transformations for $E_{st.}^{\prime i}$ and
$B_{st.}^{\prime i}$ (\ref{sk1}) are not the LT of some well-defined 4D
quantities. Therefore, contrary to the general belief, it is not true from SR
viewpoint that, e.g., [3], Jackson's Classical Electrodynamics, Sec. 11.10:
''A purely electric or magnetic field in one coordinate system will appear as
a mixture of electric and magnetic fields in another coordinate frame.''; or
that [5], Handout 10 in \textit{Physical Applications of Geometric Algebra}:
''Observers in relative motion see different fields.'' This is also exactly
proved in the tensor formalism in [13].

Both the transformations (\ref{ce}), (\ref{B}) and the transformations
(\ref{sk1}) for $E_{st.}^{\prime i}$ and $B_{st.}^{\prime i}$ (i.e., for the
3D vectors $\mathbf{E}$ and $\mathbf{B}$) are typical examples of the
`apparent' transformations that are first discussed in [10] and [11]. The
`apparent' transformations of the spatial distances (the Lorentz contraction)
and the temporal distances (the dilatation of time) are elaborated in detail
in [8] and [9] (see also [12]), and in [8] I have also discussed the
`apparent' transformations of the 3D vectors $\mathbf{E}$ and $\mathbf{B}$.
The `apparent' transformations relate, in fact, the quantities from '3+1'
space \emph{and} time (spatial and temporal distances and 3D vectors
$\mathbf{E}$ and $\mathbf{B}$) and not well-defined 4D quantities. As shown in
[8] two synchronously (for the observer) determined spatial lengths correspond
to two different 4D quantities; two temporal distances connected by the
relation for the dilatation of time also correspond to two different 4D
quantities in two relatively moving 4D inertial frames of reference, see in
[8] Figs. 3. and 4. that refer to the Lorentz contraction and the dilatation
of time respectively and compare them with Figs. 1. and 2. that refer to the
well-defined 4D quantities, the spacetime lengths for a moving rod and a
moving clock respectively. Since the spatial length, the temporal distance and
the 3D vectors $\mathbf{E}$ and $\mathbf{B}$ are different for different
observers in the 4D spacetime they do not have an independent physical
reality. It is explicitly shown in [9] that the true agreement with
experiments that test SR exists when the theory deals with well-defined 4D
quantities, i.e., the quantities that are invariant upon the passive LT; they
do not change for different observers in the 4D spacetime.

\emph{These results (both with the active and the passive LT) entail that the
standard transformations of the 3D vectors} $\mathbf{E}$ \emph{and}
$\mathbf{B}$ \emph{are not mathematically correct in the 4D spacetime, which
means that} \emph{the 3D vectors }$\mathbf{E}$ \emph{and} $\mathbf{B}$
\emph{themselves are not correctly defined quantities from the SR viewpoint}.
Consequently \emph{the usual Maxwell equations with 3D }$\mathbf{E}%
$\emph{\ and }$\mathbf{B}$\emph{\ are not in agreement with SR and they are
not physically equivalent with the relativistically correct field equations
with }$F$ (e.g., eq. (8.1) in [4], Space-Time Algebra). The same conclusion is
achieved in the tensor formalism in $\left[  13\right]  .$\emph{\bigskip}

\noindent\textit{3.3. The\ LT and the standard transformations of }%
$\mathbf{E}_{H},$ $\mathbf{B}_{H}$ \textit{and} $\mathbf{E}_{J},$
$\mathbf{B}_{J}$ \medskip

In this section, for the completness, we shall repeat the proof from Secs.
3.1. and 3.2. but using $\mathbf{E}_{H},$ $\mathbf{B}_{H}$ from [4,5] and
$\mathbf{E}_{J},$ $\mathbf{B}_{J}$ from [6]. In [4,5], as explained in Sec.
2.2., $F$ is decomposed in terms of bivectors $\mathbf{E}_{H}$ and
$\mathbf{B}_{H}$, while in [6] $F$ is decomposed in terms of 1-vector
$\mathbf{E}_{J}$ and a bivector $\mathbf{B}_{J}$. Our aim is to show that the
relativistic incorrectness of the standard transformations for the 3D vectors
$\mathbf{E}$ and $\mathbf{B}$ will be obtained regardless of the used algebric
objects for the representation of the electric and magnetic parts in the
decomposition of $F$. The correct transformations will be always, as in Sec.
3.1., simply obtained by the use of the LT of the considered 4D algebric
objects. Thus it is unimportant which algebric objects represent the electric
and magnetic fields. What is important is the way in which their
transformations are derived.

First we present this proof for $\mathbf{E}_{H},$ $\mathbf{B}_{H}$. In [4,5],
as already said in Sec. 2.2, the bivector field $F$ is expressed in terms of
the sum of a relative vector $\mathbf{E}_{H}$ and a relative bivector
$\gamma_{5}\mathbf{B}_{H}$ by making a space-time split in the $\gamma_{0}$ -
frame, eq. (\ref{FB}); $\mathbf{E}_{H}=(F\cdot\gamma_{0})\gamma_{0}$ and
$\gamma_{5}\mathbf{B}_{H}=(1/c)(F\wedge\gamma_{0})\gamma_{0}$. All these
quantities can be written as coordinate-based geometric quantities in the
standard basis $\left\{  \gamma_{\mu}\right\}  .$ Thus
\begin{equation}
\mathbf{E}_{H}=F^{i0}\gamma_{i}\wedge\gamma_{0},\quad\mathbf{B}_{H}%
=(1/2c)\varepsilon^{kli0}F_{kl}\gamma_{i}\wedge\gamma_{0}. \label{aj}%
\end{equation}
It is seen from (\ref{aj}) that both bivectors $\mathbf{E}_{H}$ and
$\mathbf{B}_{H}$ are parallel to $\gamma_{0}$, that is, it holds that
$\mathbf{E}_{H}\wedge\gamma_{0}=\mathbf{B}_{H}\wedge\gamma_{0}=0$. Further we
see from (\ref{aj}) that the components of $\mathbf{E}_{H},$ $\mathbf{B}_{H}$
in the $\left\{  \gamma_{\mu}\right\}  $ basis (i.e., in the Einstein system
of coordinates) give rise to the tensor (components)$\;(\mathbf{E}_{H}%
)^{\mu\nu}=\gamma^{\nu}\cdot(\gamma^{\mu}\cdot\mathbf{E}_{H})=(\gamma^{\nu
}\wedge\gamma^{\mu})\cdot\mathbf{E}_{H},$ (and the same for $(\mathbf{B}%
_{H})^{\mu\nu}$) which, written out as a matrix, have entries
\begin{align}
(\mathbf{E}_{H})^{i0}  &  =F^{i0}=-(\mathbf{E}_{H})^{0i}=E^{i},\quad
(\mathbf{E}_{H})^{ij}=0,\nonumber\\
(\mathbf{B}_{H})^{i0}  &  =(1/2c)\varepsilon^{kli0}F_{kl}=-(\mathbf{B}%
_{H})^{0i}=B^{i},\quad(\mathbf{B}_{H})^{ij}=0. \label{ad}%
\end{align}

Using the results from Sec. 3.1. we now apply the active LT to $\mathbf{E}%
_{H}$ and $\mathbf{B}_{H}$ from (\ref{aj}). For simplicity, as in Secs. 3.1.
and 3.2., we again consider boosts in the direction $\gamma_{1}$ for which the
rotor $R$ is given by the relation (\ref{err}) with $\gamma_{1}$ replacing
$n.$ Then using (\ref{aj}) the Lorentz transformed $\mathbf{E}_{H}^{\prime}$
can be written as
\begin{align}
\mathbf{E}_{H}^{\prime}  &  =R[(F\cdot\gamma_{0})\gamma_{0}]\widetilde
{R}=E^{1}\gamma_{1}\wedge\gamma_{0}+\gamma(E^{2}\gamma_{2}\wedge\gamma
_{0}+\nonumber\\
&  E^{3}\gamma_{3}\wedge\gamma_{0})-\beta\gamma(E^{2}\gamma_{2}\wedge
\gamma_{1}+E^{3}\gamma_{3}\wedge\gamma_{1}). \label{eh}%
\end{align}
The components $(\mathbf{E}_{H}^{\prime})^{\mu\nu}$ that are different from
zero are $(\mathbf{E}_{H}^{\prime})^{01}=-E^{1}$, $(\mathbf{E}_{H}^{\prime
})^{02}=-\gamma E^{2},$ $(\mathbf{E}_{H}^{\prime})^{03}=-\gamma E^{\text{3}},$
$(\mathbf{E}_{H}^{\prime})^{12}=\beta\gamma E^{2}$, $(\mathbf{E}_{H}^{\prime
})^{13}=\beta\gamma E^{2}$. $(\mathbf{E}_{H}^{\prime})^{\mu\nu}$ is
antisymmetric, i.e., $(\mathbf{E}_{H}^{\prime})^{\nu\mu}=-(\mathbf{E}%
_{H}^{\prime})^{\mu\nu}$ and we denoted, as in (\ref{ad}), $E^{i}=F^{i0}$.
Similarly we find for $\mathbf{B}_{H}^{\prime}$
\begin{align}
\mathbf{B}_{H}^{\prime}  &  =R[(-1/c)\gamma_{5}((F\wedge\gamma_{0})\cdot
\gamma_{0})]\widetilde{R}=B^{1}\gamma_{1}\wedge\gamma_{0}+\nonumber\\
&  \gamma(B^{2}\gamma_{2}\wedge\gamma_{0}+B^{3}\gamma_{3}\wedge\gamma
_{0})-\beta\gamma(B^{2}\gamma_{2}\wedge\gamma_{1}+B^{3}\gamma_{3}\wedge
\gamma_{1}). \label{Be}%
\end{align}
The components $(\mathbf{B}_{H}^{\prime})^{\mu\nu}$ that are different from
zero are $(\mathbf{B}_{H}^{\prime})^{01}=-B^{1}$, $(\mathbf{B}_{H}^{\prime
})^{02}=-\gamma B^{2},$ $(\mathbf{B}_{H}^{\prime})^{03}=-\gamma B^{\text{3}},$
$(\mathbf{B}_{H}^{\prime})^{12}=\beta\gamma B^{2}$, $(\mathbf{B}_{H}^{\prime
})^{13}=\beta\gamma B^{3}$. $(\mathbf{B}_{H}^{\prime})^{\mu\nu}$ is
antisymmetric, i.e., $(\mathbf{B}_{H}^{\prime})^{\nu\mu}=-(\mathbf{B}%
_{H}^{\prime})^{\mu\nu}$ and we denoted, as in (\ref{ad}), $B^{i}%
=(1/2c)\varepsilon^{kli0}F_{kl}$. Both (\ref{eh}) and (\ref{Be}) are the
familiar forms for the active LT of the bivectors, here $\mathbf{E}_{H}$ and
$\mathbf{B}_{H}$. It is important to note that $\mathbf{E}_{H}^{\prime}$ and
$\mathbf{B}_{H}^{\prime}$, in contrast to $\mathbf{E}_{H}$ and $\mathbf{B}%
_{H}$, are not parallel to $\gamma_{0}$, i.e., it \emph{does not hold} that
$\mathbf{E}_{H}^{\prime}\wedge\gamma_{0}=\mathbf{B}_{H}^{\prime}\wedge
\gamma_{0}=0$ and thus there are $(\mathbf{E}_{H}^{\prime})^{ij}\neq0$ and
$(\mathbf{B}_{H}^{\prime})^{ij}\neq0.$ Further, as in Sec. 3.1., \emph{the
components} $(\mathbf{E}_{H})^{\mu\nu}$ ($(\mathbf{B}_{H})^{\mu\nu}$)
\emph{transform upon the active LT again to the components} $(\mathbf{E}%
_{H}^{\prime})^{\mu\nu}$ ($(\mathbf{B}_{H}^{\prime})^{\mu\nu}$); \emph{there
is no mixing of components}. \emph{Thus} \emph{by the active LT}
$\mathbf{E}_{H}$ \emph{transforms to} $\mathbf{E}_{H}^{\prime}$ \emph{and}
$\mathbf{B}_{H}$ \emph{to }$\mathbf{B}_{H}^{\prime}.$ Actually, as we said,
this is the way in which every bivector transforms upon the active LT.

However \emph{the standard transformations for} $\mathbf{E}_{H,st}^{\prime}$
\emph{and} $\mathbf{B}_{H,st}^{\prime}$ \emph{are derived wrongly assuming
that the quantities obtained by the active LT of} $\mathbf{E}_{H}$ \emph{and}
$\mathbf{B}_{H}$ \emph{are again parallel to} $\gamma_{0}$\emph{, i.e., that
again holds} $\mathbf{E}_{H}^{\prime}\wedge\gamma_{0}=\mathbf{B}_{H}^{\prime
}\wedge\gamma_{0}=0$ and consequently that $(\mathbf{E}_{H,st}^{\prime}%
)^{ij}=(\mathbf{B}_{H,st}^{\prime})^{ij}=0.$ Thence, in contrast to the
correct LT of $\mathbf{E}_{H}$ \emph{and} $\mathbf{B}_{H},$ (\ref{eh}) and
(\ref{Be}) respectively, it is taken in standard derivations ([4], Space-Time
Algebra (eq. (18.22)), New Foundations for Classical Mechanics (Ch. 9 eqs.
(3.51a,b)) that
\begin{align}
\mathbf{E}_{H,st}^{\prime}  &  =[(RF\widetilde{R})\cdot\gamma_{0}]\gamma
_{0}=(F^{\prime}\cdot\gamma_{0})\gamma_{0}=E^{1}\gamma_{1}\wedge\gamma
_{0}+\nonumber\\
&  (\gamma E^{2}-\beta\gamma cB^{3})\gamma_{2}\wedge\gamma_{0}+(\gamma
E^{3}+\beta\gamma cB^{2})\gamma_{3}\wedge\gamma_{0}, \label{es}%
\end{align}
where $F^{\prime}=RF\widetilde{R}$. Similarly we find for $\mathbf{B}%
_{H,st}^{\prime}$
\begin{align}
\mathbf{B}_{H,st}^{\prime}  &  =(-1/c)\gamma_{5}[(F^{\prime}\wedge\gamma
_{0})\cdot\gamma_{0})]=B^{1}\gamma_{1}\wedge\gamma_{0}+\nonumber\\
&  (\gamma B^{2}+\beta\gamma E^{3}/c)\gamma_{2}\wedge\gamma_{0}+(\gamma
B^{3}-\beta\gamma E^{2}/c)\gamma_{3}\wedge\gamma_{0}. \label{bes}%
\end{align}
The relations (\ref{es}) and (\ref{bes}) give the familiar expressions for the
standard transformations of the 3D vectors $\mathbf{E}$ and $\mathbf{B.}$ Now,
in contrast to the correct LT of $\mathbf{E}_{H}$ \emph{and} $\mathbf{B}_{H},$
(\ref{eh}) and (\ref{Be}) respectively, \emph{the components} \emph{of the
transformed }$\mathbf{E}_{H,st}^{\prime}$ \emph{are expressed by the mixture
of components} $E^{i}$ \emph{and} $B^{i},$ \emph{and the same holds for}
$\mathbf{B}_{H,st}^{\prime}$.

The same procedure can be easily applied to the transformations of
$\mathbf{E}_{J},$ $\mathbf{B}_{J}$ from [6] and it will lead to the same
fundamental difference between the standard transformations of $\mathbf{E}%
_{J},$ $\mathbf{B}_{J}$ obtained in [6] and their correct LT. Again the active
LT of $\mathbf{E}_{J},$ $\mathbf{B}_{J}$ will be given by
\begin{equation}
\mathbf{E}_{J}^{\prime}=R(F\cdot\gamma_{0})\widetilde{R},\quad\mathbf{B}%
_{J}^{\prime}=R[-(1/c)\gamma_{5}(F\wedge\gamma_{0})]\widetilde{R},\label{jn}%
\end{equation}
while the standard transformations from [6] will follow from
\begin{equation}
\mathbf{E}_{J,st}^{\prime}=(RF\widetilde{R})\cdot\gamma_{0},\quad
\mathbf{B}_{J,st}^{\prime}=-(1/c)[\gamma_{5}((RF\widetilde{R})\wedge\gamma
_{0})].\label{jn1}%
\end{equation}
For brevity the whole discussion will not be done here. Of course the
discussion from Sec. 3.2. regarding the passive LT and the `apparent'
transformations applies in the same measure to the results of this section.
\bigskip\medskip

\noindent\textbf{4. The\ 1-vectors of\ the\ electric\ and\ magnetic\ fields\ }%
$E$ \textbf{and} $B$ \bigskip

In order to have the electric and magnetic fields defined without reference
frames, i.e., \emph{independent of the chosen reference frame and of the
chosen system of coordinates in it}, one has to replace $\gamma_{0}$ (the
velocity in units of c of an observer at rest in the $\gamma_{0}$-system) in
the relation (\ref{ebg}) (and (\ref{FB}), (\ref{J}) as well) with $v$. The
velocity $v$ and all other quantities entering into the relations (\ref{ebg})
(and (\ref{FB}), (\ref{J}) as well), but with $v$ replacing $\gamma_{0},$ are
then defined without reference frames. $v$ characterizes some general
observer. We can say, as in tensor formalism [14], that $v$ is the velocity
(1-vector) of a family of observers who measures $E$ and $B$ fields. With such
replacement the relation (\ref{ebg}) becomes
\begin{align}
F  &  =(1/c)E\wedge v+(e_{5}B)\cdot v,\nonumber\\
E  &  =(1/c)F\cdot v,\quad B=-(1/c^{2})e_{5}(F\wedge v), \label{itf}%
\end{align}
and it holds that $E\cdot v=B\cdot v=0$. Of course \emph{the relations for}
$E$ \emph{and }$B$ (\ref{itf})\emph{\ are independent of the chosen observer;
i.e., they hold for any observer.} When some reference frame is chosen with
the Einstein system of coordinates in it and when $v$ is specified to be in
the time direction in that frame, i.e., $v=c\gamma_{0}$, then all results of
the classical electromagnetism are recovered in that frame. Namely we can
always select a particular - but otherwise arbitrary - inertial frame of
reference $S,$ the frame of our 'fiducial' observers in which $v=c\gamma_{0}$
and consequently the temporal components of $E_{f}^{\mu}$ and $B_{f}^{\mu}$
are zero (the subscript '$f$' is for 'fiducial'). Then in that frame the usual
Maxwell equations for the spatial components $E_{f}^{i}$ and $B_{f}^{i}$ (of
$E_{f}^{\mu}$ and $B_{f}^{\mu}$) will be fulfilled. As a consequence the usual
Maxwell equations can explain all experiments that are performed in one
reference frame. Thus the correspondence principle is simply and naturally
satisfied. However as shown above the temporal components of $E_{f}^{\prime
\mu}$ and $B_{f}^{\prime\mu}$ are not zero; (\ref{nle}) and (\ref{nlb}) are
relativistically correct, but it is not the case with (\ref{ce}) and
(\ref{B}). This means that the usual Maxwell equations cannot be used for the
explanation of any experiment that test SR, i.e., in which relatively moving
observers have to compare their data \emph{obtained by measurements on the
same physical object.} However, in contrast to the description of the
electromagnetism with the 3D $\mathbf{E}$ and $\mathbf{B,}$ the description
with $E$ and $B$ is correct not only in that frame but in all other relatively
moving frames and it holds for any permissible choice of coordinates. It is
worth noting that the relations (\ref{itf}) are not the definitions of $E$ and
$B$ but they are the relations that connect two equivalent formulations of
electrodynamics, the standard formulations with the $F$ field and the new one
with the $E$ and $B$ fields. Every of these formulations is an independent,
complete and consistent formulation. For more detail see [7] where four
equivalent formulations are presented, the $F$ and $E,$ $B$ - formulations and
two new additional formulations with real and complex combinations of $E$ and
$B$ fields. All four formulations are given in terms of quantities that are
defined without reference frames. In the recent work [15] I have presented the
formulation of relativistic electrodynamics (independent of the reference
frame and of the chosen system of coordinates in it) that uses the bivector
field $F.$ This formulation with $F$ field is, as already said, a
self-contained, complete and consistent formulation that dispenses with either
electric and magnetic fields or the electromagnetic potentials. Note however
that in the $E,$ $B$ - formulation of electrodynamics in [7] the expression
for the stress-energy vector $T(v)$ and all quantities derived from $T(v)$ are
written for the special case when $v,$ the velocity of observers who measure
$E$ and $B$ fields is $v=cn$, where $n$ is the unit normal to a hypersurface
through which the flow of energy-momentum ($T(n)$) is calculated. The more
general case with $v\neq n$ will be reported elsewhere.

In addition, as we have already said, the replacement of $\gamma_{0}$ with $v$
in the relations (\ref{FB}) and (\ref{J}) also yields the electric and
magnetic fields defined without reference frames. However, it is much simpler
and, in fact, closer to the classical formulation of the electromagnetism with
the 3D $\mathbf{E}$ and $\mathbf{B}$ to work with 1-vectors $E$ and $B$
instead of to use the bivectors $\mathbf{E}_{H}$, $\mathbf{B}_{H}$ and
$\mathbf{B}_{J}$ (but all with $v$ replacing $\gamma_{0}$).

We have not mentioned some other references that refer to the Clifford algebra
formalism and its application to the electrodynamics as are, e.g., [16]. The
reason is that they use the Clifford algebra formalism with spinors but, of
course, they also erroneously consider that the standard transformations of
the 3D $\mathbf{E}$ and $\mathbf{B}$ (\ref{sk1}) are the LT of the electric
and magnetic fields. \bigskip\medskip

\noindent\textbf{5. Conclusions}\bigskip

The whole consideration explicitly shows that the 3D quantities $\mathbf{E}$
and $\mathbf{B}$, their transformations and the equations with them are
ill-defined in the 4D spacetime. More generally, \emph{the 3D quantities do
not have an independent physical reality in the 4D spacetime. }Contrary to the
general belief we find that \emph{it is not true from the SR viewpoint that
observers in relative motion see different fields; the transformations}
(\ref{ce}), (\ref{B}) \emph{and} (\ref{sk1}) \emph{(or }(\ref{es})\emph{ and
}(\ref{bes}\emph{)) are not relativistically correct. According to the
relativistically correct transformations, the LT} (\ref{nle}) \emph{and}
(\ref{nlb}), \emph{(or }(\ref{eh})\emph{ and }(\ref{Be}\emph{))} \emph{the
electric field transforms only} \emph{to the electric field and the same holds
for the magnetic field.} Thence the relativistically correct physics must be
formulated with 4D quantities that are defined without reference frames, or by
the 4D coordinate-based geometric quantities, e.g., as in [7] in the Clifford
algebra formalism with multivectors, or [8,9] in the tensor formalism. The
principle of relativity is automatically included in such theory with
well-defined 4D quantities, while in the standard approach to SR [1] it is
postulated outside the mathematical formulation of the theory. The comparison
with experiments from [9] (and [7]) reveals that the true agreement with
experiments that test SR can be achieved when such well-defined 4D quantities
are considered.\bigskip\medskip

\noindent\textbf{Acknowledgment \bigskip}

I am grateful to Professor Larry Horwitz for his continuos interest, support
and useful comments in the development of my invariant SR. \bigskip\medskip

\noindent\textbf{References}\bigskip

\noindent$\left[  1\right]  $ A. Einstein, Ann. Physik. 17 (1905) 891, tr. by
W. Perrett and G.B.

Jeffery, in The Principle of Relativity, Dover, New York.

\noindent$\left[  2\right]  $ H.A. Lorentz, Proceedings of the Academy of
Sciences of Amsterdam

(1904) 6, in W. Perrett and G.B. Jeffery, The Principle of Relativity,

Dover, New York.

\noindent$\left[  3\right]  $ J.D. Jackson, Classical Electrodynamics, Wiley,
New York, 1977 2nd

edn.; L.D. Landau and E.M. Lifshitz, The Classical Theory of Fields,

Pergamon, Oxford, 1979 4th edn.; C.W. Misner, K.S.Thorne, and J.A.

Wheeler, Gravitation, Freeman, San Francisco, 1970.

\noindent$\left[  4\right]  $ D. Hestenes, Space-Time Algebra, Gordon and
Breach, New York, 1966;

Space-Time Calculus, available at: http://modelingnts.la. asu.edu/evolution.

html; New Foundations for Classical Mechanics, Kluwer Academic

Publishers, Dordrecht, 1999 2nd. edn..

\noindent$\left[  5\right]  $ S. Gull, C. Doran, and A. Lasenby, in W.E.
Baylis (Ed.), Clifford

(Geometric) Algebras with Applications to Physics, Mathematics,

and Engineering, Birkhauser, Boston, 1997 Chs. 6-8.; C. Doran, and

A. Lasenby, Physical Applications of Geometric Algebra, available at:

www.mrao.cam. ac.uk/\symbol{126}Clifford/

\noindent$\left[  6\right]  $ B. Jancewicz, Multivectors and Clifford Algebra
in Electrodynamics,

World Scientific, Singapore, 1989.

\noindent$\left[  7\right]  $ T. Ivezi\'{c}, hep-th/0207250; hep-ph/0205277.

\noindent$\left[  8\right]  $ T. Ivezi\'{c}, Found. Phys. 31 (2001) 1139.

\noindent$\left[  9\right]  $ T. Ivezi\'{c}, Found. Phys. Lett. 15 (2002) 27;
physics/0103026; physics/0101091.

\noindent$\left[  10\right]  $ F. Rohrlich, Nuovo Cimento B 45 (1966) 76.

\noindent$\left[  11\right]  $ A. Gamba, Am. J. Phys. 35 (1967) 83.

\noindent$\left[  12\right]  $ T. Ivezi\'{c}, Found. Phys. Lett. 12 (1999)
105; Found. Phys. Lett. 12

(1999) 507.

\noindent$\left[  13\right]  $ T. Ivezi\'{c}, hep-th/0302188; to be published
in Found. Phys. 33 (2003)

(issue 9).

\noindent$\left[  14\right]  $ R.M. Wald, General Relativity, The University
of Chicago Press, Chicago,

1984.

\noindent$\left[  15\right]  $ T. Ivezi\'{c}, physics/0305092.

\noindent$\left[  16\right]  $ W.E. Baylis, Electrodynamics, a Modern
Geometric Approach, Birkh\"{a}user,

Boston, 1998; P. Lounesto, Clifford Algebras and Spinors, Cambridge

University, Cambridge, 1997.
\end{document}